\documentclass{article} % For LaTeX2e
\usepackage{iclr2027_conference,times}

\usepackage{amsmath,amsfonts,bm}

\def\eqref#1{equation~\ref{#1}}
\def\1{\bm{1}}

\DeclareMathAlphabet{\mathsfit}{\encodingdefault}{\sfdefault}{m}{sl}
\SetMathAlphabet{\mathsfit}{bold}{\encodingdefault}{\sfdefault}{bx}{n}

\usepackage{hyperref}
\usepackage{url}
\usepackage{booktabs}
\usepackage{tabularx}
\usepackage{array}
\usepackage{makecell}
\usepackage{multirow}
\usepackage{graphicx}

\usepackage{longtable}
\usepackage{ragged2e}
\usepackage{pdflscape}

\newcolumntype{P}[1]{>{\RaggedRight\arraybackslash}p{#1}}

\title{RECOB: Reliable Benchmarking of Experimental Optimization in Chemistry and Materials Science}

\author{
Zikai Xie$^{1,2\dagger}$\thanks{Corresponding authors.}, Jiaming Wan$^{3\dagger}$, Linjiang Chen$^{1,2,4*}$ \\
$^{1}$State Key Laboratory of Precision and Intelligent Chemistry, University of Science and Technology of China\\
$^{2}$Center for Scientific Intelligence Innovation\\
$^{3}$Department of Computer Science, Fudan University \\
$^{4}$School of Chemistry, School of Computer Science, University of Birmingham \\
$^{\dagger}$Equal Contribution. \\
}

\iclrfinalcopy % Uncomment for camera-ready version, but NOT for submission.
\begin{document}

\maketitle

\begin{abstract}
Optimization methods for experimental science are often evaluated on
synthetic functions that are reproducible but omit important
characteristics of real experiments. We introduce RECOB (REliable Chem
Optimization Benchmark, Github repository: \hyperlink{https://github.com/XieZikai/RECOB}{https://github.com/XieZikai/RECOB}
), a black-box optimization benchmark constructed
exclusively from data generated through physical experiments in chemistry
and materials science. The suite contains 14 single-objective and two
multi-objective tasks spanning chemical reactions, material formulations,
electrochemical systems, continuous-flow processes, and automated
laboratories. Each task provides a machine-readable specification of its
decision variables, feasible domain, physical constraints, objective
direction, and experimental provenance. Continuously queryable learned
oracles are screened using repeated holdout validation and prespecified
admission criteria, while measured-table replay enables evaluation using
the original experimental responses. Under a common paired evaluation protocol, we compare ten single-objective
and eight multi-objective optimization methods. HEBO achieves the best
aggregate single-objective rank, while qNEHVI leads the multi-objective
comparison. Model-based methods generally outperform non-adaptive
baselines, although their computational overhead varies substantially.
We further assess benchmark reliability using independent oracle retraining and measured-table replay. Aggregate optimizer rankings remain highly
consistent across retrained oracles, while replay preserves the broad
performance hierarchy using only physically measured responses. Together,
these results show that RECOB can reproducibly distinguish optimizer
performance as an experimentally grounded and reliability-tested
benchmark for black-box optimization in chemistry and materials science.
\end{abstract}

\section{Introduction}
\label{sec:intro}

Many discovery and process-development problems in experimental science involve optimizing experimental conditions, material compositions, or hyperparameter tuning~\citep{yu2026optimizing,luo2025physics,wu2019hyperparameter}. These tasks can often be formulated as expensive black-box optimization problems: the objective has no analytical expression, gradients are inaccessible, and each evaluation requires a physical experiment~\citep{jones1998efficient}. Because experiments are typically time-consuming and may consume costly materials, the central goal is to identify high-performing conditions as rapidly as possible within a limited evaluation budget~\citep{shields2021bayesian}. Bayesian optimization (BO), evolutionary algorithms (EA), and other derivative-free methods have therefore been widely adopted for closed-loop experimental design in chemistry, materials science, and automated laboratories~\citep{shahriari2015taking,back1993overview,burger2020mobile,dave2022autonomous}.

Developing and comparing these methods requires testbeds that are reproducible, inexpensive to evaluate, and representative of real applications. Analytical test functions and standardized suites such as COCO/BBOB support exact evaluation and large-scale replication, but often fail to capture mixed variable types, physical constraints, and compositional domains~\citep{hansen2021coco,hansen2009real}. Recent benchmarks have narrowed this gap using experimental reaction tables~\citep{shields2021bayesian}, pool-based materials datasets~\citep{liang2021benchmarking}, mechanistic simulators~\citep{felton2021summit}, and learned emulators~\citep{hase2021olympus}. However, three important challenges remain. First, publicly available tasks cover a limited range of experimental systems, and their constraints, conditional spaces, and provenance are rarely represented in a unified, machine-readable format~\citep{hickman2022bayesian}. Second, prior work on learned benchmarks has shown that predictive quality alone is insufficient to establish benchmark fidelity: downstream optimizer rankings and search trajectories must also be preserved~\citep{zela2020surrogate,pfisterer2022yahpo}. Third, finite candidate tables and continuously queryable emulators define different optimization problems, yet their agreement has not been systematically studied.

To address these challenges, we introduce \textbf{RECOB} (\textbf{RE}liable \textbf{C}hem \textbf{O}ptimization \textbf{B}enchmark), a black-box optimization benchmark constructed from real chemistry and materials experiments. We curate 19 experimental datasets originating from approximately 17 peer-reviewed publications. All response values are grounded in physical experiments rather than analytical test functions or computational-property calculations. Our inclusion criteria require that the input variables be experimentally controllable, that the objectives be directly measured or have a clearly defined experimental interpretation, and that data collection in the original study be explicitly associated with experimental optimization or design-space exploration. For each dataset, we reconstruct the variable types, experimental bounds, joint constraints, conditional spaces, objective directions, campaign information, and data provenance.

We train AutoGluon ensembles as continuously queryable oracles and screen them using five repeated holdout splits and prespecified admission criteria~\citep{erickson2020autogluon}. Based on predictive accuracy, worst-split performance, and between-split variability, tasks are assigned to core or conditional tiers. For single-objective evaluation, we select 14 tasks from nine independent publications and compare ten optimizers under a paired protocol in which all algorithms share the same initial design for each task--trial pair. We assess final and anytime performance, cross-task ranking stability, and candidate-generation cost. For datasets with multiple scientifically meaningful objectives, we also construct a multi-objective extension covering hypervolume-based, scalarization-based, information-theoretic, and evolutionary methods.

We evaluate benchmark robustness in two ways. First, we repeat the benchmark using independently trained oracles with different training seeds to quantify the stability of optimizer rankings. Second, we introduce a measured-candidate track that returns recorded responses from the original experimental tables. We then compare optimizer rankings between the two protocols.

Our main contributions are as follows:
\begin{enumerate}
    \item We construct RECOB, an experimentally grounded benchmark comprising 14 single-objective and two multi-objective tasks across real-world problems in chemistry and materials science. Each task includes machine-readable variable definitions, feasible domains, physical constraints, objective directions, and data provenance.

    \item We establish reproducible continuous-oracle and measured-candidate evaluation protocols, and compare ten single-objective and eight multi-objective optimizers under shared initial designs and equal evaluation budgets. The evaluation covers final and anytime performance, cross-task rankings, and optimizer computational overhead.

    \item We assess benchmark reliability through independently trained oracles and measured-table replay. Optimizer rankings are highly stable across oracle training seeds, while replay preserves broad performance groups but not exact rankings on every task.
\end{enumerate}

\section{Related Work}
\label{sec:related}

Optimization benchmarks range from analytical test suites to domain-specific virtual environments. COCO/BBOB~\citep{hansen2021coco,hansen2009real} provides exactly and repeatedly evaluable functions for large-scale algorithm comparison, but does not reproduce many features of experimental search spaces, such as mixed variables, compositional constraints, and historically sampled domains. Summit~\citep{felton2021summit} improves domain relevance through chemistry-motivated mechanistic and data-driven virtual environments, while remaining primarily simulator- or emulator-based.

Benchmarks constructed from historical experiments generally follow either a finite-candidate or a learned-emulator protocol. Shields et al.~\citep{shields2021bayesian} evaluated reaction-optimization methods on high-throughput experimental tables, while Liang et al.~\citep{liang2021benchmarking} used pool-based campaigns across several materials datasets. These protocols preserve recorded experimental responses but restrict evaluation to previously measured candidates. Olympus~\citep{hase2021olympus, hickman2023olympus}, by contrast, trained continuously queryable emulators from experimental datasets, enabling optimization beyond the original candidate tables but introducing dependence on learned response models.

Validation of learned benchmarks has been studied more explicitly in other domains. YAHPO Gym~\citep{pfisterer2022yahpo} showed that predictive metrics should be complemented by downstream checks of optimization behavior and optimizer rankings. Following this principle, RECOB screens oracle predictive quality before task admission and evaluates whether aggregate optimizer conclusions remain stable across independently trained oracles and measured-table replay. Table~\ref{tab:benchmark_comparison} summarizes the resulting differences in data provenance, evaluation protocol, validation, and task support.

\begin{table*}[th!]
    \centering
    \caption{
        Comparison of representative black-box optimization benchmarks.
    }
    \label{tab:benchmark_comparison}

    \scriptsize
    \setlength{\tabcolsep}{2.5pt}
    \renewcommand{\arraystretch}{1.15}

    \begin{tabularx}{\linewidth}{
        @{}
        >{\raggedright\arraybackslash}p{2.25cm}
        *{4}{>{\centering\arraybackslash}X}
        >{\centering\arraybackslash}p{2.45cm}
        *{2}{>{\centering\arraybackslash}X}
        @{}
    }
        \toprule

        \multirow{2}{*}{\textbf{Benchmark}}
        &
        \multicolumn{2}{c}{\textbf{Data}}
        &
        \multicolumn{3}{c}{\textbf{Oracle and validation}}
        &
        \multicolumn{2}{c}{\textbf{Task support}}
        \\

        \cmidrule(lr){2-3}
        \cmidrule(lr){4-6}
        \cmidrule(lr){7-8}

        &
        \makecell{\textbf{Physical}\\\textbf{experiments}}
        &
        \makecell{\textbf{Measured}\\\textbf{track}}
        &
        \makecell{\textbf{Continuous}\\\textbf{oracle}}
        &
        \makecell{\textbf{Predictive}\\\textbf{validation}}
        &
        \makecell{\textbf{Rank/trajectory}\\\textbf{audit}}
        &
        \makecell{\textbf{Constraints}\\\textbf{/ mixed}}
        &
        \textbf{Objectives}
        \\

        \midrule

        COCO/BBOB
        &
        No
        &
        No
        &
        Analytic
        &
        N/A
        &
        N/A
        &
        Limited
        &
        SO + MO
        \\

        Summit
        &
        Partial
        &
        Not std.
        &
        \makecell{Simulator /\\emulator}
        &
        Task-specific
        &
        Not systematic
        &
        Yes
        &
        SO + MO
        \\

        Shields et al.
        &
        Yes
        &
        Yes
        &
        No
        &
        \makecell{N/A}
        &
        \makecell{N/A}
        &
        \makecell{Primarily\\discrete}
        &
        Primarily SO
        \\

        Liang et al.
        &
        Yes
        &
        Yes
        &
        No
        &
        \makecell{N/A}
        &
        \makecell{N/A}
        &
        Limited
        &
        SO
        \\

        Olympus
        &
        Yes
        &
        Not std.
        &
        Learned
        &
        Yes
        &
        Not systematic
        &
        Yes
        &
        \makecell{SO +\\limited MO}
        \\

        YAHPO Gym
        &
        No
        &
        \makecell{Tabular\\comparison}
        &
        Learned
        &
        Yes
        &
        Yes
        &
        Yes
        &
        SO + MO
        \\

        \textbf{RECOB}
        &
        \textbf{Yes}
        &
        \textbf{Yes}
        &
        \textbf{Learned}
        &
        \textbf{Yes}
        &
        \textbf{Yes}
        &
        \textbf{Yes}
        &
        \textbf{SO + MO}
        \\

        \bottomrule
    \end{tabularx}

    \begin{minipage}{0.98\textwidth}
        \footnotesize
        \textit{Notes.}
        ``Measured track'' denotes an explicitly defined evaluation protocol
        based on previously measured candidates.
        ``Not std.'' means that relevant data may be available, but the
        corresponding track is not standardized as part of the benchmark.
        SO and MO denote single- and multi-objective optimization,
        respectively.
    \end{minipage}
\end{table*}

\begin{table}[th!]
    \centering
    \caption{Frozen quantitative criteria for learned-oracle admission.}
    \label{tab:oracle_tiers}
    \small
    \begin{tabular}{lccc}
        \toprule
        Criterion
        & \makecell{Core\\(all required)}
        & \makecell{Conditional\\(all required)}
        & \makecell{Not admitted\\(any sufficient)} \\
        \midrule
        Labeled observations
        & $\geq 50$
        & $\geq 40$
        & $<40$ \\

        Splits with $R^2>0$
        & $\geq 4/5$
        & $\geq 3/5$
        & $<3/5$ \\

        Mean $R^2$
        & $\geq 0.50$
        & $\geq 0.20$
        & $<0.20$ \\

        Mean Pearson $r$
        & $\geq 0.70$
        & $\geq 0.50$
        & $<0.50$ \\

        Mean range-NRMSE
        & $\leq 0.15$
        & $\leq 0.25$
        & $>0.25$ \\
        \bottomrule
    \end{tabular}

    \vspace{2pt}
\end{table}

\section{Benchmark construction}
\subsection{Data Processing}

RECOB was designed to represent optimization in experimentally measured systems. We therefore collected data only from source studies reporting physical experiments, in which controllable experimental variables and measured responses could be formulated as an optimization problem. Experimental tables were obtained from the article text, Supporting Information, author-maintained repositories, or previously released benchmark datasets. The complete list of source studies and provenance records is provided in Appendix~\ref{app:a} Table~\ref{tab:paper_license_source}.

For each dataset, variables explicitly treated as optimization decisions in the source study were used as input features, while experimentally measured quantities identified as optimization objectives were used as response variables. Variable names and units were standardized without altering the measured values. No missing-value imputation or general outlier filtering was applied to the model-ready data.

Each dataset was then represented by a machine-readable optimization specification defining the external decision variables, model inputs, variable types and units, allowed categories or numerical bounds, joint and nonlinear constraints, and objective direction. When a source contained chemically distinct reactions or experimental campaigns, these were represented as separate optimization tasks.

The initial collection comprised 19 experimental datasets from 17 source studies, containing 5,291 experimental records and 43 candidate target--space combinations. These candidates were subsequently evaluated through the oracle-construction and screening procedure described below.

\subsection{Oracle Construction}

Continuously queryable oracles were constructed using AutoGluon Tabular~\citep{erickson2020autogluon}. A common automated machine-learning pipeline was used across all datasets to accommodate differences in sample size, dimensionality, scale, and variable type, while reducing task-specific discretion in model and hyperparameter selection. Before fitting the frozen oracle on all available data, we evaluated out-of-sample predictive quality using five fixed 80/20 holdout splits. Candidate tasks with insufficient or unstable validation performance were excluded from the continuous-oracle benchmark, thereby limiting the suite to tasks for which the experimentally measured input--response relationship could be reproduced with acceptable accuracy. For candidate tasks containing repeated exact input configurations, all repetitions were assigned to the same partition, and the resulting group-split metrics were used for screening to prevent leakage between training and validation data.

For each split, we computed the coefficient of determination ($R^2$), Pearson correlation coefficient ($r$), and range-normalized root mean squared error,
\[
\mathrm{NRMSE}_{\mathrm{range}}
=
\frac{\sqrt{n^{-1}\sum_i (y_i-\hat{y}_i)^2}}
     {y_{\max}-y_{\min}},
\]
where $y_{\max}$ and $y_{\min}$ denote the maximum and minimum measured responses for the corresponding task. Oracle admission was jointly determined by the number of labeled observations, the mean $R^2$, mean Pearson $r$, mean range-NRMSE, and the number of validation splits with positive $R^2$.

\begin{table*}[th]
\centering
\caption{Predictive validation results and admission outcomes for the
14 retained single-objective tasks.}
\label{tab:oracle_screening_results}
\scriptsize
\setlength{\tabcolsep}{3pt}

\begin{tabularx}{\linewidth}{
    @{}
    >{\raggedright\arraybackslash}p{2.8cm}
    >{\raggedright\arraybackslash}X
    c
    c
    c
    c
    c
    c
    @{}
}
    \toprule
    Experimental source
    & Opt target
    & Reference
    & \makecell{Labeled\\observations}
    & \makecell{Positive-$R^2$\\splits}
    & \makecell{Mean\\$R^2$}
    & \makecell{Mean\\Pearson $r$}
    & \makecell{Mean\\range-NRMSE} \\
    \midrule

    AgNP synthesis
    & Reaction rate
    & \cite{low2024evolution}
    & 144 & 5/5 & 0.938 & 0.970 & 0.077 \\

    OPV formulation
    & Degradation
    & \cite{langner2020beyond}
    & 2,322 & 5/5 & 0.839 & 0.918 & 0.057 \\

    Mobile robotic chemist
    & Hydrogen evolution
    & \cite{burger2020mobile}
    & 595 & 5/5 & 0.821 & 0.913 & 0.103 \\

    BRINE
    & Conductivity, campaign 2
    & \cite{ramezani2026brine}
    & 121 & 5/5 & 0.763 & 0.884 & 0.114 \\

    Amide coupling
    & Yield, reaction 1
    & \cite{wagner2024self}
    & 117 & 5/5 & 0.754 & 0.886 & 0.094 \\

    3D-printing materials
    & Compression modulus
    & \cite{erps2021accelerated}
    & 143 & 5/5 & 0.718 & 0.868 & 0.083 \\

    Continuous-flow chemistry
    & N-benzylation impurity
    & \cite{schweidtmann2018machine}
    & 73 & 5/5 & 0.852 & 0.934 & 0.106 \\

    Continuous-flow chemistry
    & SNAr E-factor
    & \cite{schweidtmann2018machine}
    & 66 & 5/5 & 0.838 & 0.918 & 0.108 \\

    P3HT--CNT composites
    & Electrical conductivity$^\dagger$
    & \cite{bash2021multi}
    & 233 & 5/5 & 0.694 & 0.840 & 0.113 \\

    3D-printing materials
    & Toughness
    & \cite{erps2021accelerated}
    & 143 & 5/5 & 0.517 & 0.740 & 0.112 \\

    \midrule

    Li-ion electrolyte
    & Conductivity$^\dagger$
    & \cite{dave2022autonomous}
    & 125 & 4/5 & 0.376 & 0.618 & 0.157 \\

    Amide coupling
    & Yield, reaction 4
    & \cite{wagner2024self}
    & 40 & 5/5 & 0.529 & 0.832 & 0.231 \\

    Amide coupling
    & Yield, reaction 2
    & \cite{wagner2024self}
    & 72 & 5/5 & 0.454 & 0.711 & 0.177 \\

    BRINE
    & Conductivity, campaign 1
    & \cite{ramezani2026brine}
    & 114 & 5/5 & 0.261 & 0.621 & 0.147 \\

    \bottomrule
\end{tabularx}

\begin{minipage}{0.98\linewidth}
    \footnotesize
    \textit{Notes.}
    Rows above and below the intermediate horizontal rule correspond
    to Core and Conditional tasks, respectively.
    Tasks marked with $^\dagger$ were evaluated using exact-input
    group splits; all other tasks used five repeated random 80/20
    holdout splits. Classification follows the criteria in
    Table~\ref{tab:oracle_tiers}.
\end{minipage}
\end{table*}

The quantitative criteria used for oracle admission are summarized in Table~\ref{tab:oracle_tiers}. A candidate was assigned to the Core set only if it satisfied all Core criteria. Among the remaining candidates, those satisfying all Conditional criteria were assigned to the Conditional set, whereas failure of any Conditional criterion resulted in exclusion from the aggregate continuous-oracle evaluation. Quantitative screening was applied only after the task definition and feasible optimization domain had been established during data processing.

%这里需要一个表，把筛选过后的任务的来源、文章（只放引用即可，不放名字）、每个指标、最终判别为什么类别都记录下来

Application of these criteria produced 10 Core and four Conditional tasks, of which the results are reported in Table~\ref{tab:oracle_screening_results}. Across the 14 admitted tasks, mean $R^2$ ranged from 0.261 to 0.938, mean Pearson correlation from 0.621 to 0.970, and mean range-NRMSE from 0.057 to 0.231. The tier assignments are provided as task-level quality metadata rather than treated as an experimental factor in the optimizer comparison.

Following screening, the selected AutoGluon ensemble for each task was refitted using all valid labeled observations and frozen as the deployed oracle. Experimental noise was not explicitly simulated because its sources and distributions are task-specific and cannot be faithfully represented by a common Gaussian noise model. Every candidate is checked against the declared schema and feasibility constraints before prediction; invalid categories, off-grid values, non-finite inputs, and bound or joint-constraint violations are rejected rather than silently corrected. Multi-objective oracles are constructed by evaluating compatible single-objective models on the same validated candidate and concatenating their predictions, without fitting additional models. Training configurations, validation results, artifact hashes, and per-task screening decisions are provided in the Supporting Information.

\section{Evaluation Protocol}

\subsection{Benchmark Interaction and Fairness Controls}

RECOB follows a sequential ask--evaluate--tell protocol. At iteration $t$, an optimizer proposes a candidate $\mathbf{x}_t$ for the task domain $\mathcal{X}$. After validation, the corresponding model features are passed to the frozen learned oracle, which returns a scalar response $\hat{y}_t=\hat{f}(\mathbf{x}_t)$ for a single-objective task, or a response vector $\hat{\mathbf{y}}_t=[\hat{f}_1(\mathbf{x}_t),\ldots,\hat{f}_m(\mathbf{x}_t)]$ for a multi-objective task. Each response is supplied to the optimizer before its next proposal, giving a sequential protocol with batch size 1. Multi-objective tasks reuse the corresponding frozen single-output oracles and do not require an additional multi-output model.

The benchmark interface supports continuous, discrete, categorical, and mixed-variable search spaces. Directly submitted candidates are checked against the bounds, legal categories, step grids, and joint or nonlinear constraints defined by the task. In particular, a step-constrained variable is valid only when $x_j=l_j+k\Delta_j$ for some $k\in\mathbb{Z}$, within the stored numerical tolerance. In the formal algorithm comparisons, optimizer outputs were first converted into legal external candidates using the same task-aware domain mappings, including grid mapping for step-constrained variables, and were then passed through the common validator. Derived model inputs were computed only after validation. Thus, all algorithms were evaluated over the same declared feasible domain.

We used a paired design comprising 20 trials per task. Each task--trial pair was associated with a frozen initial design containing ten feasible, non-duplicate conditions. Every algorithm received the same 10 conditions in the same order and observed the corresponding oracle responses before algorithm-specific sampling began. Post-initialization random streams were generated using a deterministic seed policy, with independent optimizer seeds derived for each task, trial, and algorithm.

Each run contained ten initialization evaluations followed by 100 sequential optimization evaluations, giving a total budget of
$N_{\mathrm{eval}}=10+100=110$. Internal model fitting and acquisition optimization did not receive additional oracle calls. In the single-objective experiments, duplicate decoded proposals were resampled before evaluation, with the associated suggestion time retained; in the multi-objective experiments, repeated candidates remained in the recorded evaluation history and budget. No per-suggestion performance timeout was imposed in the completed formal experiments, and checkpointing allowed interrupted runs to resume without altering their histories. For a given task, all algorithms used the same oracle artifact, task definition, validation logic, objective direction, and initial-design assets. 

Optimization performance and computational overhead were reported separately: the single-objective study records candidate-suggestion time, whereas the multi-objective runs retain end-to-end run time. All formal runs were submitted through Slurm to a dual-socket compute node equipped with two AMD EPYC 7702 64-core processors and 256~GB of DDR4 memory. Each task--algorithm--trial run was scheduled as an independent job with two CPU cores and 8~GB of memory. Reported suggestion times include only the recorded candidate-generation time of completed runs and exclude Slurm queueing time.

\subsection{Evaluated Optimizers and Performance Metrics}

The evaluated algorithms were selected to represent a broad range of search
mechanisms. We considered only methods that can operate without
task-specific scientific prior knowledge; each optimizer received only the
variable types, feasible domain, constraints, and observations provided by
the benchmark. The ten single-objective methods comprise two non-adaptive
baselines (Random Search and scrambled Sobol Search~\citep{owen1998scrambling});
three Gaussian-process BO methods (BoTorch LogEI, BoTorch
UCB~\citep{balandat2020botorch}, and BoFire
LogEI~\citep{durholt2025bofire}); a SMAC-inspired random-forest
expected-improvement method~\citep{hutter2011sequential}; a
density-estimation method (Optuna TPE~\citep{akiba2019optuna}); an
evolution strategy (CMA-ES~\citep{hansen2003reducing}); and two Bayesian
optimization frameworks designed for complex search spaces
(HEBO~\citep{cowen2022hebo} and Gryffin~\citep{hase2021gryffin}).
The RF/SMAC-like method is our own random-forest ensemble implementation
and does not use the official SMAC software.

The eight multi-objective methods comprise Random Search and Sobol Search
as non-adaptive baselines; NSGA-II~\citep{deb2002fast} as an evolutionary
method; qNEHVI~\citep{daulton2021parallel} as a hypervolume-based Bayesian
optimization method; qNParEGO~\citep{daulton2021parallel}
as a scalarization-based method; TSEMO~\citep{bradford2018efficient} as a
Thompson-sampling method; and MESMO~\citep{belakaria2019max} and
JES~\citep{tu2022joint} as information-theoretic methods. The qNEHVI and
qNParEGO experiments use their numerically stable logarithmic
implementations. All multi-objective methods propose one new candidate per
iteration. Implementation details and hyperparameters of all single- and multi-objective methods are provided in
Appendix~\ref{app:b}.

\paragraph{Agentic Methods.}
LLM-based agents are increasingly being explored for data analysis,
scientific modeling, and iterative optimization
~\citep{hu2024infiagent,zhao2026scienceflow}. However, their numerical
reasoning remains brittle~\citep{li2025exposing,yan2025large}. We therefore focus
the main evaluation on conventional black-box optimization algorithms rather
than attempting a comprehensive comparison of agentic systems. As a targeted
preliminary case study, we evaluate ScienceFlow on the single-objective tasks
and report the setup and results in Appendix~\ref{app:agentic}.

For single-objective tasks, minimization and maximization directions were first converted to a common maximization convention. Performance was quantified using normalized best-so-far simple regret. For each task \(d\), we pooled the objective values from the frozen union of all formal evaluation histories across all algorithms and trials. Let \(z_d^{\mathrm{best}}\) and \(z_d^{\mathrm{worst}}\) denote, respectively, the best and worst values in this pooled set under the maximization convention, which were shared by all algorithms evaluated on the same task. If \(b_t\) denotes the best value observed by evaluation \(t\), the normalized simple regret is
\(\tilde{r}_t=
\max\left\{0,\,
\frac{z_d^{\mathrm{best}}-b_t}
{z_d^{\mathrm{best}}-z_d^{\mathrm{worst}}}
\right\}.\)
Final optimization performance was measured by \(\tilde{r}_{110}\), while optimization speed was measured by the normalized trapezoidal area under the best-so-far regret curve from the end of the common initialization (evaluation 10) through evaluation 110. Lower values are better for both metrics. Means and standard deviations were calculated over the 280 task–trial runs for each algorithm; because every task contributed the same number of trials, this aggregation assigned equal weight to the 14 tasks. For each task, algorithms were ranked according to their median final normalized simple regret over the 20 trials, with average ranks and fractional wins assigned to ties; the overall rank was the mean of these within-task ranks.

For multi-objective tasks, performance was evaluated using the hypervolume
of the nondominated set. Within each task and trial, every objective was
normalized using the range of the ten common initial observations, and the
dominated reference point was fixed at $(-0.1,\ldots,-0.1)$ in maximization
space. Because this normalization is applied to the objectives rather than
to the hypervolume itself, the resulting HV is not bounded by one. Final
Pareto-front quality was measured by the HV after 110 evaluations, while
anytime performance was measured by the normalized area under the
hypervolume curve (AUC-HV) from the end of the common initialization to
evaluation 110. Larger values are better for both metrics. For each task
and algorithm, final HV and AUC-HV were summarized as the mean and standard
deviation over the 20 paired trials. Algorithms were ranked by final HV
within each task--trial block, and the reported mean rank was averaged
across the 20 trials and two tasks; AUC-HV was not used in this ranking.

\section{Experimental Results}

\subsection{Benchmark Results}

We evaluated ten single-objective optimizers on the 14 RECOB tasks.
Table~\ref{tab:single_objective_ranking} summarizes final and anytime
performance. HEBO achieved the best final aggregate performance, with the
lowest mean task rank and the lowest mean final normalized simple regret.
It won eight tasks outright and shared the best median result on two
others. BoFire LogEI and BoTorch LogEI occupied the next two positions,
followed by a more closely grouped set of RF/SMAC-like EI, BoTorch UCB, and
CMA-ES. Gryffin, Sobol Search, and Random Search consistently remained near
the bottom of the final ranking. The overall differences were significant according to the Friedman test
($\chi^2=81.30$, $p=8.92\times10^{-14}$). Pairwise Wilcoxon
signed-rank tests indicated that
HEBO significantly outperformed all other methods except BoFire LogEI
($p=0.084$) and BoTorch LogEI ($p=0.084$).

Final performance and optimization speed did not produce exactly the same
ordering. BoTorch LogEI obtained the lowest normalized regret AUC,
indicating the strongest overall early-budget performance, while BoFire
LogEI and HEBO followed closely. HEBO continued to improve over the longer
budget and ultimately achieved the strongest final rank. CMA-ES and
RF/SMAC-like EI also improved progressively, whereas Random and Sobol Search
fell behind because their proposals did not adapt to the observed
responses.

As a sensitivity analysis, we recomputed the single-objective rankings separately for the two fidelity tiers. The two rankings
were broadly consistent (Spearman $\rho=0.794$): HEBO ranked first in
both subsets, whereas Gryffin, Sobol Search, and Random Search remained
in the lower-performing group. Conditional tasks
showed lower mean normalized final regret and regret AUC than the Core
tasks, although neither difference was statistically significant at the
task level. Thus, the admission tier did not introduce a qualitative
change in the main performance hierarchy.

Performance remained task dependent. GP-based methods, CMA-ES, and TPE
each won at least one problem, while HEBO was strongest across a broader
subset of the suite. Final discrimination
was weaker for OPV degradation and printing compression modulus because
several algorithms reached the same oracle plateaus, although their regret
AUC values and optimization trajectories still distinguished how rapidly
those plateaus were reached. 

\paragraph{Computational overhead.}

The mean suggestion time varied by more than five orders of magnitude.
HEBO required approximately 3.50 seconds per optimization step, compared
with 16.91 and 14.65 seconds for BoFire LogEI and BoTorch LogEI,
respectively. RF/SMAC-like EI retained competitive performance with only
0.17 seconds per step, providing a favorable performance--cost compromise,
while CMA-ES and Optuna TPE incurred negligible suggestion overhead.
Gryffin had both the largest mean suggestion time and a low final rank.
These measurements describe successful candidate-generation steps and
were not incorporated into the performance ranking.

Eight multi-objective methods were evaluated on the three-objective amide
task and the two-objective printing task. As shown in
Table~\ref{tab:multiobjective_ranking}, qNEHVI achieved the best overall
final-HV rank and the highest final HV on the amide task, whereas MESMO
obtained the highest final HV on printing. qNEHVI also achieved the highest
mean AUC-HV on both tasks, indicating that it constructed useful
nondominated sets more rapidly across the full evaluation budget.

On the amide task, qNEHVI ranked first in 18 of 20 paired trials and
significantly outperformed every alternative. Its
performance reflected a favorable balance across the three reaction
objectives rather than improvement in only one response. On printing,
MESMO, qNEHVI, and qNParEGO formed the leading group, and their final-HV
differences were not statistically decisive after multiple-comparison
correction.

Computational requirements differed substantially among the
multi-objective methods. qNParEGO had the lowest total wall time among the
three leading model-based approaches, while retaining the third-best
overall rank. NSGA-II required only 0.33 cumulative hours and improved
substantially over Random and Sobol Search, although it did not match the
final or anytime performance of the Bayesian methods. TSEMO incurred the
largest total wall time but ranked fifth.

Overall, the multi-objective results distinguish adaptive from
non-adaptive strategies as well as different Bayesian and evolutionary
search mechanisms. Nevertheless, the cross-task ranking remains
descriptive because the current multi-objective suite contains only two
continuous, all-maximization tasks, and HV values are comparable only
within the same task.

\begin{table*}[t]
    \centering
    \caption{Aggregate single-objective performance across 14 tasks.
    Regret values are reported as mean $\pm$ standard deviation over
    280 task--trial runs. Mean suggestion time excludes the ten shared
    initialization evaluations.}
    \label{tab:single_objective_ranking}
    \small
    \setlength{\tabcolsep}{4pt}
    \begin{tabular}{lccccc}
        \toprule
        Algorithm
        & \makecell{Mean\\rank}
        & \makecell{Task\\wins}
        & \makecell{Final normalized\\simple regret $\downarrow$}
        & \makecell{Normalized regret\\AUC $\downarrow$}
        & \makecell{Mean suggestion time\\(s step$^{-1}$) $\downarrow$} \\
        \midrule
        HEBO
        & 1.71 & 8.5
        & $0.062 \pm 0.117$
        & $0.126 \pm 0.148$
        & 3.50 \\

        BoFire LogEI
        & 3.96 & 1.5
        & $0.075 \pm 0.114$
        & $0.125 \pm 0.145$
        & 16.91 \\

        BoTorch LogEI
        & 4.29 & 1.5
        & $0.071 \pm 0.112$
        & $0.111 \pm 0.133$
        & 14.65 \\

        RF/SMAC-like EI
        & 4.43 & 0
        & $0.080 \pm 0.080$
        & $0.146 \pm 0.130$
        & 0.170 \\

        BoTorch UCB
        & 4.46 & 0.5
        & $0.115 \pm 0.169$
        & $0.149 \pm 0.183$
        & 12.00 \\

        CMA-ES
        & 4.57 & 1.0
        & $0.084 \pm 0.089$
        & $0.160 \pm 0.118$
        & $5.88\times10^{-4}$ \\

        Optuna TPE
        & 5.43 & 1.0
        & $0.105 \pm 0.108$
        & $0.172 \pm 0.142$
        & $9.54\times10^{-3}$ \\

        Gryffin
        & 8.36 & 0
        & $0.178 \pm 0.146$
        & $0.231 \pm 0.156$
        & 34.29 \\

        Sobol Search
        & 8.50 & 0
        & $0.189 \pm 0.144$
        & $0.230 \pm 0.151$
        & $8.47\times10^{-5}$ \\

        Random Search
        & 9.29 & 0
        & $0.200 \pm 0.154$
        & $0.240 \pm 0.161$
        & $5.71\times10^{-5}$ \\
        \bottomrule
    \end{tabular}
\end{table*}

\begin{table*}[t]
    \centering
    \caption{Final and anytime multi-objective performance over 20 paired
    trials per task. HV and AUC-HV are reported as mean $\pm$ standard
    deviation. Total wall time is summed over the 40 successful runs of
    each algorithm and excludes scheduler queueing and unsuccessful
    preliminary attempts.}
    \label{tab:multiobjective_ranking}
    \small
    \setlength{\tabcolsep}{3.5pt}
    \begin{tabular}{lcccccc}
        \toprule
        Algorithm
        & \makecell{Amide\\final HV $\uparrow$}
        & \makecell{Amide\\AUC-HV $\uparrow$}
        & \makecell{Printing\\final HV $\uparrow$}
        & \makecell{Printing\\AUC-HV $\uparrow$}
        & \makecell{Final-HV\\mean rank}
        & \makecell{Total wall\\time (h) $\downarrow$} \\
        \midrule
        qNEHVI
        & $3.048 \pm 0.809$
        & $2.835 \pm 0.724$
        & $2.619 \pm 0.568$
        & $2.301 \pm 0.475$
        & 1.675 & 22.29 \\

        MESMO
        & $2.907 \pm 0.749$
        & $2.768 \pm 0.701$
        & $2.648 \pm 0.606$
        & $2.289 \pm 0.487$
        & 2.200 & 17.39 \\

        qNParEGO
        & $2.851 \pm 0.769$
        & $2.579 \pm 0.657$
        & $2.591 \pm 0.597$
        & $2.251 \pm 0.477$
        & 3.050 & 11.79 \\

        JES
        & $2.865 \pm 0.757$
        & $2.700 \pm 0.702$
        & $2.434 \pm 0.613$
        & $2.023 \pm 0.456$
        & 3.425 & 18.04 \\

        TSEMO
        & $2.728 \pm 0.690$
        & $2.450 \pm 0.574$
        & $2.199 \pm 0.473$
        & $1.790 \pm 0.374$
        & 4.825 & 27.06 \\

        NSGA-II
        & $2.347 \pm 0.552$
        & $1.876 \pm 0.432$
        & $1.906 \pm 0.501$
        & $1.521 \pm 0.324$
        & 6.075 & 0.33 \\

        Sobol Search
        & $2.036 \pm 0.503$
        & $1.750 \pm 0.360$
        & $1.606 \pm 0.371$
        & $1.399 \pm 0.275$
        & 7.275 & 0.41 \\

        Random Search
        & $1.876 \pm 0.606$
        & $1.652 \pm 0.603$
        & $1.580 \pm 0.389$
        & $1.387 \pm 0.301$
        & 7.475 & 0.38 \\
        \bottomrule
    \end{tabular}
\end{table*}

\subsection{Benchmark Fidelity and Robustness}

We evaluated RECOB reliability through oracle retraining and
measured-table replay. The former tests whether benchmark conclusions are
sensitive to randomness in oracle construction, whereas the latter
provides a comparison using the original measured
experimental responses. Additional quantitative results for these analyses, including prediction agreement across independent oracle fits, replay-pool construction, and task-level matching-distance diagnostics, are provided in Appendix~\ref{app:d}.

For the retraining analysis, independent AutoGluon oracles were trained
with seeds 2026, 2027, and 2028 for six representative tasks: AgNP
reaction rate, amide reaction-1 yield, BRINE campaign-2 conductivity,
mobile hydrogen evolution, OPV degradation, and printing compression
modulus. Pairwise correlations between the resulting ten-algorithm
rankings were consistently high, with Spearman correlations of
0.964--1.000 and Kendall correlations of 0.911--1.000. The leading, intermediate, and lower-performing groups were preserved
across all three retrainings: HEBO, BoFire LogEI, and BoTorch LogEI
consistently occupied the leading group; BoTorch UCB, RF/SMAC-like EI,
CMA-ES, and Optuna TPE remained in the intermediate group; and Gryffin,
Sobol Search, and Random Search consistently occupied the final three
positions. These results show that the optimizer comparisons are
insensitive to the random seed used to train the AutoGluon oracles.

The numerical predictions were similarly stable when independently
trained oracles were evaluated on shared response-independent candidate
sets. Pairwise prediction correlations were 0.986--0.999, rank
correlations were at least 0.976, and range-normalized mean absolute
errors were 0.0046--0.0179. Thus, the independently trained oracles also produced closely agreeing predictions.

For measured-table replay, numerical variables were scaled by their declared ranges, while categorical mismatches were assigned a distance of one and matches a distance of zero. Each proposal was matched to the nearest unused experimental row. The matched candidate and its recorded response were returned to the optimizer, after which the row was removed from the candidate pool. At the matched budget of 60 evaluations,
the learned-oracle and replay rankings had a Spearman correlation of
0.758 and a Kendall correlation of 0.556. BoFire LogEI and BoTorch LogEI
remained among the top three methods under both protocols, while BoTorch
UCB obtained the best replay rank. HEBO moved from the leading
learned-oracle group to sixth place under replay. At the lower end,
Gryffin, Sobol Search, and Random Search continued to occupy the final
three positions. Therefore, replay preserved the broad distinction between the
stronger model-based methods and the lower-performing baselines, despite
some changes within the leading and intermediate groups.

At 110 evaluations, the measured-table replay produced a lower aggregate
rank correlation of $\rho=0.455$ because of the
finite-candidate evaluation mechanism. Each optimizer proposal is mapped
to the nearest unused experimental row, which is subsequently removed
from the candidate pool. As the number of replay iterations increases,
different algorithms increasingly encounter overlapping high-performing
records, while candidate depletion and final-value ties compress their
performance differences. Consequently, measured-table replay becomes
less discriminative at larger evaluation budgets, and small differences
in the order in which rows are selected can produce larger changes in the
final ranks.

HEBO showed the largest change between the two protocols. Its strong
continuous-oracle performance suggests that it benefits from concentrating
search within narrow promising regions of a continuously queryable
response surface. Under replay, however, each proposal is replaced by the
nearest unused experimental row. Once nearby rows have been evaluated,
subsequent proposals may be matched to increasingly different conditions,
weakening the advantage of concentrated continuous search. 

The two analyses revealed different levels of sensitivity: oracle retraining had little effect on optimizer rankings, whereas measured-table replay changed the ordering within the leading group. Independent AutoGluon retraining produced highly consistent oracle predictions and algorithm rankings, indicating that the aggregate conclusions were not driven by a particular oracle fit. Measured-table replay, in contrast, retained the broad separation between stronger model-based methods and lower-performing baselines using only experimentally recorded responses, although the ordering within the leading group changed. RECOB therefore supports reproducible aggregate comparisons, while exact optimizer rankings should be interpreted in the context of the evaluation protocol.

\section{Limitations and Conclusion}

In this work, we introduced RECOB, an experimentally grounded benchmark
for black-box optimization in chemistry and materials science. Across 14
single-objective and two multi-objective tasks, RECOB revealed clear but
task-dependent differences among optimization strategies. HEBO achieved the best aggregate single-objective rank among the evaluated methods, while qNEHVI achieved the best aggregate rank on the two evaluated multi-objective tasks. Optimizer overhead varied by more than orders of magnitude, showing a trade-off between optimization performance and calculation wall-time.

Optimizer rankings were highly consistent across independently retrained
learned oracles, and measured-table replay preserved the broad performance
hierarchy using physically measured responses. These results provide strong evidence that RECOB can reliably distinguish broad performance differences among black-box optimization algorithms. RECOB remains limited by the coverage and sampling density of its source experiments, does not explicitly model experimental noise or replicate variability, and cannot replace prospective laboratory validation. Within these limits, RECOB enables controlled comparison of optimization algorithms on experimentally derived problems before their evaluation in prospective laboratory campaigns.

\subsection*{Acknowledgement}

Zikai Xie gratefully acknowledge the National Advanced Talent Cultivation Center for Chemistry, USTC, for its support.

\subsection*{AI use statement}

In this work, we used generative AI tools for assisting with translation, cleaning and reformatting datasets, supporting qualitative and thematic data analysis or interpreting results. We have not used generative AI tools for generating synthetic data sets or conceptual frameworks, designing or providing feedback on research methodology or experiments,
and the rest of the required disclosure tasks are not applicable to this work.
Additionally, we used generative AI tools for creating and editing software code, and editing the paper to improve readability. We have reviewed all AI-assisted work. We checked that the datasets reformatted by the LLM were consistent with the original datasets and that the statistical analysis tools provided by the LLM were correct. We take responsibility for the final content of this work,
including text, claims or artifacts produced with the aid of generative AI.

\subsection*{Ethics statement}

This work exclusively uses publicly available data. All third-party data were accessed and used in accordance with the licenses specified by their original sources. Detailed information on the data sources and the licenses under which the data were made available and used in this study is provided in Appendix Table~\ref{tab:paper_license_source}.

\subsection*{Reproducibility statement}

The supplementary repository contains the processed datasets, machine-readable
task and constraint specifications, frozen learned-oracle manifests, common
initial designs, raw evaluation histories, and analysis scripts used in this
study. Dataset provenance and licensing are listed in Appendix~\ref{app:a}.
Oracle fitting, domain mappings, seed construction, optimizer hyperparameters,
and software environments are documented in Appendix~\ref{app:b}. The exact
single-objective normalization endpoints and task-level results are reported in
Appendix~\ref{app:c}, and the oracle-retraining and measured-table replay
protocols are detailed in Appendix~\ref{app:d}. Released data, configurations,
and initial-design assets are accompanied by SHA-256 checksums so that the
inputs to a reproduced run can be verified independently.

%\subsubsection*{Author Contributions}
%If you'd like to, you may include  a section for author contributions as is done in many journals. This is optional and at the discretion of the authors.

%\subsubsection*{Acknowledgments}
%Use unnumbered third level headings for the acknowledgments. Allacknowledgments, including those to funding agencies, go at the end of the paper.

\bibliography{iclr2027_conference}
\bibliographystyle{iclr2027_conference}

\appendix
\section{Data source and licensing}
\label{app:a}

\begin{landscape}
\begin{longtable}{
    @{}
    P{0.8cm}
    P{16.2cm}
    P{2.3cm}
    P{3.6cm}
    @{}
}
\caption{Summary of papers, licenses, and sources}
\label{tab:paper_license_source}\\

\toprule
\textbf{No.} & \textbf{Paper} & \textbf{License} & \textbf{Source} \\
\midrule
\endfirsthead

\multicolumn{4}{c}{\tablename~\thetable{} (continued)}\\
\toprule
\textbf{No.} & \textbf{Paper} & \textbf{License} & \textbf{Source} \\
\midrule
\endhead

\midrule
\multicolumn{4}{r}{Continued on next page}\\
\endfoot

\bottomrule
\endlastfoot

1 &
An Integrated Self-Optimizing Programmable Chemical Synthesis and Reaction Engine~\citep{leonov2024integrated} &
CC BY 4.0 &
Nature Communications \\

2 &
Bayesian Self-Optimization for Telescoped Continuous Flow Synthesis~\citep{clayton2023bayesian} &
CC BY 4.0 &
Angewandte Chemie \\

3 &
Rapid and Mild One-Flow Synthetic Approach to Unsymmetrical Sulfamides Guided by Bayesian Optimization~\citep{sugisawa2021rapid} &
GPL-2.0 &
GitHub \\

4 &
A Mobile Robotic Chemist~\citep{burger2020mobile} &
Apache License &
GitHub \\

5 &
A Self-Driving Laboratory Advances the Pareto Front for Material Properties~\citep{macleod2022self} &
CC BY 4.0 &
Nature Communications \\

6 &
A Slug Flow Platform with Multiple Process Analytics Facilitates Flexible Reaction Optimization~\citep{wagner2024slug} &
CC BY 4.0 &
Advanced Science \\

7 &
Accelerated Discovery of 3D Printing Materials Using Data-Driven Multiobjective Optimization~\citep{erps2021accelerated} &
CC BY-NC 4.0 &
Science Advances \\

8 &
Autonomous Optimization of Non-Aqueous Li-Ion Battery Electrolytes via Robotic Experimentation and Machine Learning Coupling~\citep{dave2022autonomous} &
CC BY 4.0 &
Nature Communications \\

9 &
Beyond Ternary OPV: High-Throughput Experimentation and Self-Driving Laboratories Optimize Multicomponent Systems~\citep{langner2020beyond} &
CC BY-NC 4.0 &
Advanced Materials \\

10 &
BRINE: A Cost-Effective Electrochemical Self-Driving Laboratory for Accelerated Discovery of High-Performance Electrolytes~\citep{ramezani2026brine} &
CC BY 3.0 &
Digital Discovery \\

11 &
Data-Driven Product-Process Optimization of N-Isopropylacrylamide Microgel Flow-Synthesis~\citep{kaven2024data} &
CC BY 4.0 &
arXiv \\

12 &
Evolution-Guided Bayesian Optimization for Constrained Multi-Objective Optimization in Self-Driving Labs~\citep{low2024evolution} &
CC BY 4.0 &
npj Computational Materials \\

13 &
Machine Learning Meets Continuous Flow Chemistry: Automated Optimization Towards the Pareto Front of Multiple Objectives~\citep{schweidtmann2018machine} &
CC BY 4.0 &
Chemical Engineering Journal \\

14 &
Multi-Fidelity High-Throughput Optimization of Electrical Conductivity in P3HT-CNT Composites~\citep{bash2021multi} &
MIT License &
GitHub \\

15 &
Self-Optimizing Flow Reactions for Sustainability: An Experimental Bayesian Optimization Study~\citep{wagner2024self} &
CC BY 4.0 &
ACS Sustainable Chemistry \& Engineering \\

16 &
Toward Practical Design of High-Entropy Catalysts for Chlorine Evolution Reaction via Pareto-Guided Multi-Objective Bayesian Optimization Enabled by a Robotic AI-Chemist~\citep{yang2026toward} &
MIT License &
GitHub \\

17 &
Physics-Informed, Dual-Objective Optimization of High-Entropy-Alloy Nanozymes by a Robotic AI Chemist~\citep{luo2025physics} &
MIT License &
GitHub \\

\end{longtable}

\noindent\footnotesize
\textit{License abbreviations:}
CC BY = Creative Commons Attribution;
CC BY-NC = Creative Commons Attribution--NonCommercial.

\end{landscape}

\normalsize
\section{Implementation details}
\label{app:b}

\subsection{Oracle screening, fitting, and release artifacts}

Candidate targets were evaluated on five train--test splits with seeds 42--46.
Random 80/20 splits were used except for the Li-ion electrolyte and P3HT--CNT
tasks, for which rows sharing the same input vector were kept in the same split.
The $R^2$, Pearson correlation, and range-normalized RMSE values in
Table~\ref{tab:oracle_screening_results} are the means over these five splits;
the positive-split count records the number of splits with $R^2>0$. Screening
was performed before the final oracle was fitted, so the formal optimization
histories were not used in task admission or model selection.

The same final fitting protocol was used for every admitted target. AutoGluon
TabularPredictor was run with \texttt{medium\_quality}, a 15-s fitting budget,
root-mean-squared error as the evaluation metric, two CPU cores, sequential
fitting, and candidate model families \texttt{RF} and \texttt{XT}. The selected
\texttt{WeightedEnsemble\_L2} was then refitted on all valid labeled rows using
\texttt{refit\_full}; the deployed model name is
\texttt{WeightedEnsemble\_L2\_FULL} for all 14 tasks. Predictions are point
predictions and no stochastic observation-noise term is added. For a
multi-objective query, the relevant frozen single-output predictors receive the
same validated candidate and their outputs are concatenated in the declared
objective order.

Each task release includes a copy of its task configuration, source metadata
and constraints, the pre-refit AutoGluon leaderboard, smoke-test predictions,
and a \texttt{training\_manifest.json}. The manifest records the training-row
count, input columns, fitting configuration, selected and deployed model names,
software versions, and full SHA-256 hashes of the processed data, metadata, and
constraint files. Candidate-level validation results are retained in
\texttt{repeated\_split\_results.csv} and
\texttt{repeated\_model\_screening.csv}; the revalidated task decisions are
retained in \texttt{latest\_model\_screening.csv}. These machine-readable files
provide the unrounded values underlying Table~\ref{tab:oracle_screening_results}.

\subsection{Task parameterization and feasible-domain mappings}

For the formal comparison, every task was represented by a latent unit cube
$[0,1]^d$ and a deterministic task-specific decoder. This provided every
optimizer with the same feasible parameterization, irrespective of whether the
optimizer library natively implements categorical, compositional, or joint
constraints. Table~\ref{tab:task_domain_mapping} summarizes the external and
latent dimensions. Complete bounds, units, category levels, step sizes, and
constraint tolerances are stored in the released \texttt{tasks.json} and
\texttt{multi\_tasks.json} files.

\begin{landscape}
\begin{longtable}{@{}P{5.0cm}P{1.4cm}P{1.7cm}P{1.5cm}P{3.7cm}P{8.6cm}@{}}
\caption{Task domains and the common feasible parameterizations used in the
formal experiments. $p$ is the number of externally submitted variables and
$d$ is the dimension of the latent unit cube.}
\label{tab:task_domain_mapping}\\
\toprule
Task & Tier & $p$ & $d$ & Variable types & Decoder and feasibility conditions \\
\midrule
\endfirsthead
\multicolumn{6}{c}{\tablename~\thetable{} (continued)}\\
\toprule
Task & Tier & $p$ & $d$ & Variable types & Decoder and feasibility conditions \\
\midrule
\endhead
\midrule
\multicolumn{6}{r}{Continued on next page}\\
\endfoot
\bottomrule
\endlastfoot
AgNP reaction rate & Core & 5 & 5 & Continuous & Box scaling followed by validation of\newline
$\mathrm{AgNO_3}/\mathrm{AA}\geq0.3$ and
$\mathrm{AgNO_3}/\mathrm{AA}+\mathrm{seed}/\mathrm{AgNO_3}\geq2$. \\
OPV degradation & Core & 5 & 5 & One categorical; four continuous & The first latent coordinate selects PBQ-QF or PTB7-Th;\newline four exponential weights are normalized to a composition summing to one. \\
Mobile hydrogen evolution & Core & 10 & 11 & Step-grid numeric & P10-MIX1 is mapped to a 0.2-mg grid.\newline Nine liquid additions are mapped to 0.25-mL grids with total volume at most 5~mL and L-cysteine at least 0.25~mL;\newline water is derived to complete 5~mL. \\
BRINE campaign 2 & Core & 5 & 6 & Continuous & Bounded composition with total dispensed volume\newline at most 330~$\mu$L. \\
Amide yield, reaction 1 & Core & 5 & 5 & Continuous & Independent box bounds. \\
Printing compression modulus & Core & 6 & 6 & Continuous composition & Bounded simplex with the six ink fractions\newline summing to 100~wt\%. \\
N-benzylation impurity & Core & 4 & 4 & Continuous & Independent box bounds. \\
SNAr E-factor & Core & 4 & 4 & Continuous & Independent box bounds. \\
P3HT--CNT conductivity & Core & 5 & 5 & Continuous composition & Bounded simplex with the five component fractions\newline summing to 100~wt\%. \\
Printing toughness & Core & 6 & 6 & Continuous composition & Bounded simplex with the six ink fractions\newline summing to 100~wt\%. \\
Li-ion electrolyte conductivity & Conditional & 3 & 3 & Continuous & Independent box bounds. \\
Amide yield, reaction 4 & Conditional & 5 & 5 & Continuous & Independent box bounds. \\
Amide yield, reaction 2 & Conditional & 5 & 5 & Continuous & Independent box bounds. \\
BRINE campaign 1 & Conditional & 4 & 5 & Continuous & Bounded composition with total dispensed volume\newline at most 330~$\mu$L. \\
\midrule
Amide reactions 1/2/4 & MO & 5 & 5 & Continuous & Intersection of the three child-task box domains;\newline three maximization objectives. \\
Printing mechanical properties & MO & 6 & 6 & Continuous composition & Six ink fractions sum to 100~wt\%;\newline compression modulus and toughness are maximized. \\
\end{longtable}
\end{landscape}

For an unconstrained numerical variable, decoding is linear between its stored
lower and upper bounds. Categorical values occupy equal intervals of a latent
coordinate. For a sum-equality constraint, positive weights
$w_j=-\log u_j$ are normalized and then projected onto the bounded simplex.
For a sum-at-most constraint over $k$ variables, one latent coordinate controls
the used fraction of the available total through $u_0^{1/k}$ and $k$ further
coordinates allocate that amount using normalized exponential weights. This
adds one latent radial coordinate, explaining $d=p+1$ for the BRINE tasks.
Step-constrained outputs are mapped to their stored grids. The special mobile
chemist decoder implements its solid and liquid grids before deriving water.
An AgNP proposal that fails either nonlinear ratio constraint is replaced by a
legal draw from the same seeded domain stream. Every decoded proposal is passed
through the released task validator before oracle evaluation.

\subsection{Initial designs, seeds, and repeated proposals}

The single-objective release contains 280 frozen initialization files: 14 tasks
by 20 trials, each with ten feasible conditions. They were generated using
task-aware scrambled Sobol sampling. Within-task duplicates were forbidden both
within and across trials, step grids and joint constraints were checked after a
CSV round trip, and every OPV trial contains five conditions from each
categorical level. For a fixed task and trial, all ten algorithms read the same
file in the same row order.

Post-initialization optimizer seeds were constructed as
\[
s=\operatorname{uint32}\!\left(
\operatorname{SHA256}(\texttt{chembench-evaluation-stream-v1}\,|\,
\texttt{task}\,|\,\texttt{trial}\,|\,\texttt{algorithm})_{1:4}
\right),
\]
where the first four digest bytes were interpreted in big-endian order. Thus,
Random Search and Sobol Search start new streams rather than continuing the
stream used to construct the common initialization. The multi-objective runner
contains 40 frozen Latin-hypercube initialization files (two tasks by 20
trials). It uses an analogous SHA-256 construction with base seed 20260820 and
stores the resulting seed in each run summary.

For single-objective runs, candidate identity uses exact categorical strings
and numerical values rounded to 12 decimal places. A repeated decoded proposal
is resampled, while all time spent on rejected suggestions remains in the
recorded suggestion time. After 100 repeated suggestions, a deterministic legal
fallback stream derived from the run seed and iteration is used; the optimizer
is told the decoded condition that was actually evaluated. For multi-objective
runs, repeated proposals are evaluated and remain part of the 110-evaluation
budget.

\subsection{Single-objective optimizer configurations}

All single-objective optimizers received the ten common observations before
their first formal suggestion. Minimization responses were sign-flipped inside
the optimizer interface, so every method operated under a maximization
convention. Unless specified below, library defaults were retained.

\begin{table*}[h]
\centering
\caption{Single-objective implementations and fixed hyperparameters.}
\label{tab:so_hyperparameters}
\footnotesize
\begin{tabularx}{\textwidth}{@{}P{3.0cm}X@{}}
\toprule
Method & Implementation used in the formal benchmark \\
\midrule
Random Search & Independent NumPy uniform draw in $[0,1]^d$ at each step. \\
Sobol Search & SciPy scrambled Sobol sequence with the run-specific seed. \\
RF/SMAC-like EI & Random forest with 128 trees, \texttt{max\_features}=0.8, and seed $s+n$ after $n$ observations. EI is evaluated on 4,096 uniform candidates using the across-tree mean and standard deviation and an exploration offset $0.01\,\mathrm{sd}(y)$. This is an in-house SMAC-inspired implementation, not the official SMAC software. \\
BoTorch LogEI & Double-precision SingleTaskGP with standardized outcomes and exact marginal-likelihood fitting; LogExpectedImprovement with current best response; acquisition optimization with $q=1$, eight restarts, and 128 raw samples. \\
BoTorch UCB & The same GP and acquisition optimization as BoTorch LogEI, with UpperConfidenceBound and $\beta=0.2$. \\
BoFire LogEI & BoFire single-objective Bayesian strategy (SoboStrategy) on $[0,1]^d$ with qLogEI and the run-specific seed; all common initial observations are supplied through the BoFire tell interface. \\
Optuna TPE & TPESampler with the run-specific seed and \texttt{n\_startup\_trials}=0; the ten common observations are inserted as completed trials. \\
CMA-ES & Initial mean equal to the mean latent coordinate of the ten common conditions; initial step size 0.25, bounds $[0,1]^d$, population size 5, and run-specific seed. \\
HEBO & Numerical HEBO design space on $[0,1]^d$, \texttt{rand\_sample}=0, and \texttt{scramble\_seed} equal to the run seed; the common observations are loaded before the first suggestion. \\
Gryffin & Continuous latent parameters on $[0,1]^d$; one CPU, \texttt{boosted}=false, 64 random samples, 200 epochs, 128 draws, learning rate 0.05, and sampling strategy 1. \\
\bottomrule
\end{tabularx}
\end{table*}

\subsection{Multi-objective optimizer configurations}

The two multi-objective tasks contain only continuous external variables, with
the printing compositions decoded onto the same bounded simplex for every
method. For qNEHVI, qNParEGO, TSEMO, MESMO, and JES, a double-precision
SingleTaskGP with a vector-valued standardized outcome transform was refitted by
exact marginal likelihood after every observation.

\begin{table*}[h]
\centering
\caption{Multi-objective implementations and fixed hyperparameters.}
\label{tab:mo_hyperparameters}
\footnotesize
\begin{tabularx}{\textwidth}{@{}P{2.8cm}X@{}}
\toprule
Method & Implementation used in the formal benchmark \\
\midrule
Random Search & Independent uniform draws in the latent unit cube. \\
Sobol Search & Scrambled Sobol sequence with the algorithm seed and a 1,024-point fast-forward before the first formal suggestion. \\
NSGA-II & pymoo NSGA-II with population size 10, the common initialization as the starting population, duplicate elimination enabled, and termination at 110 oracle evaluations. \\
qNEHVI & BoTorch qLogNoisyExpectedHypervolumeImprovement. The acquisition reference is the componentwise minimum of the ten initial responses minus 0.1 times their componentwise range. Baseline pruning is enabled; $q=1$, five restarts, and 128 raw samples are used. \\
qNParEGO & BoTorch qLogNParEGO with a new seeded nonnegative scalarization vector normalized to sum to one at every step; baseline pruning, $q=1$, five restarts, and 128 raw samples. \\
TSEMO & One joint posterior sample is drawn on 4,096 scrambled-Sobol candidates; a point is sampled from the nondominated set of that posterior draw. \\
MESMO & Eight posterior Pareto samples with ten requested Pareto points, population 512, and ten search attempts; qLowerBoundMultiObjectiveMaxValueEntropySearch with 64 acquisition samples. \\
JES & The same Pareto-sampling configuration as MESMO; qLowerBoundMultiObjectiveJointEntropySearch with 64 acquisition samples. \\
\bottomrule
\end{tabularx}
\end{table*}

MESMO and JES acquisition optimization used $q=1$, five restarts, 128 raw
samples, batch limit 1, initialization batch limit 16, and at most 200
iterations. If a posterior path contained fewer than ten detected Pareto
points, only that path was redrawn; the requested front width was reduced
through 10, 8, 6, 4, 2, and 1 only when necessary, with population 1,024 and 20
search attempts for replacement draws. Numerical acquisition failures used, in
order, ten deterministic Sobol starting points, the best finite acquisition
value among 2,048 Sobol candidates, and finally the point of maximum finite
posterior variance. Every fallback event is recorded in its run summary.

An audit of the 35,200 multi-objective evaluations found four repeated
candidates (three from qNParEGO and one from JES), corresponding to 0.011\% of
the recorded budget. The NSGA-II interface reconstructed its starting
population through the latent simplex representation; the maximum coordinate
difference from the stored initialization was $2.13\times10^{-14}$. All initial
oracle responses were identical except for one printing response, for which a
tree-model threshold changed the predicted toughness by 43.87~Pa (0.067\%).
The formal analysis uses the responses actually recorded in each history.

For multi-objective inference, algorithms were ranked by final HV within each
of the 20 paired task--trial blocks. Within each task, the method with the
largest mean final HV was compared with each of the other seven methods using a
one-sided paired Wilcoxon signed-rank test; the seven resulting $p$-values were
Holm-adjusted within that task.

\subsection{Software and compute environment}

The frozen single-objective optimizer environment used Python 3.10 and included
BoFire 0.3.1, BoTorch 0.16.1, CMA 3.2.2, HEBO 0.3.6, Optuna 4.9.0,
scikit-learn 1.7.2, SciPy 1.12.0, pandas 2.3.3, NumPy 1.24.4, pymoo 0.6.0,
and CPU PyTorch 2.13.0. Gryffin was installed from the bundled source revision
\texttt{fb149d18f9c81b96179cc682c46e05db60bc7229}. The isolated oracle
environment used AutoGluon 1.5.0, NumPy 2.2.6, pandas 2.3.3, SciPy 1.15.3,
and scikit-learn 1.7.2. The multi-objective bundle fixed NumPy 1.26.4 and
pymoo 0.6.1.5 and used BoTorch 0.16.1. The released server bundles include the
environment specifications; the single-objective and oracle environments also
include full transitive dependency locks.

Each formal task--algorithm--trial run was an independent Slurm job assigned two
CPU cores and 8~GB RAM on a dual-socket AMD EPYC 7702 node with 256~GB DDR4.
The learned oracle ran in a separate process from the optimizer environment.
Single-objective suggestion time was measured around optimizer candidate
generation only; oracle evaluation and scheduler queueing were excluded.
Multi-objective wall time covers the complete run process but excludes scheduler
queueing.

\section{Single-objective normalization and task-level results}
\label{app:c}

\subsection{Frozen normalization endpoints}

Table~\ref{tab:single_objective_normalizers} gives the exact endpoints used in
the normalized-regret definition in the main text. They were computed once from the frozen union of all 2,800 formal
single-objective histories (10 algorithms, 14 tasks, and 20 trials), after
converting minimization tasks to maximization by multiplying their responses by
$-1$. These are post hoc scale factors only: they were not supplied to an
optimizer and could not affect a candidate trajectory. Negative transformed
values for the minimization tasks therefore reflect the sign conversion rather
than negative regret.

\begin{table}[h]
\centering
\caption{Per-task pooled-history endpoints under the common maximization
convention. The denominator of normalized regret is
$z_d^{\mathrm{best}}-z_d^{\mathrm{worst}}$.}
\label{tab:single_objective_normalizers}
\small
\begin{tabular}{lcrr}
\toprule
Task & Original direction & $z_d^{\mathrm{best}}$ & $z_d^{\mathrm{worst}}$ \\
\midrule
AgNP reaction rate & Max & 0.047508 & 0.004150 \\
OPV degradation & Min & 0.015795 & $-0.709751$ \\
Mobile hydrogen evolution & Max & 19.982903 & 0.200546 \\
BRINE campaign 2 & Max & 25.925722 & 11.595119 \\
Amide yield, reaction 1 & Max & 97.753334 & 21.826666 \\
Printing compression modulus & Max & 2.760167 & 0.233667 \\
N-benzylation impurity & Min & $-2.291167$ & $-9.816167$ \\
SNAr E-factor & Min & $-0.247800$ & $-1.919633$ \\
P3HT--CNT conductivity & Max & 742.199402 & 9.429895 \\
Printing toughness & Max & 97438.640625 & 38700.308594 \\
Li-ion electrolyte conductivity & Max & 13.827521 & 4.084336 \\
Amide yield, reaction 4 & Max & 48.230000 & 0.803333 \\
Amide yield, reaction 2 & Max & 98.386665 & 51.549999 \\
BRINE campaign 1 & Max & 28.430731 & 15.822099 \\
\bottomrule
\end{tabular}
\end{table}

\subsection{Per-task final performance and tier sensitivity}

Table~\ref{tab:task_level_single_objective} reports the task-level medians that
were ranked to produce Table~\ref{tab:single_objective_ranking}. Exact zeroes
indicate that at least half of the trials reached the pooled empirical best;
values printed as $<0.001$ are positive. The complete-precision values and all
20 trial-level histories are included in the release.

\begin{landscape}
\begin{longtable}{@{}P{4.6cm}rrrrrrrrrr@{}}
\caption{Median final normalized simple regret over 20 trials for each
single-objective task. Lower is better. BT-L, RF, BT-U, and RS denote BoTorch
LogEI, RF/SMAC-like EI, BoTorch UCB, and Random Search, respectively.}
\label{tab:task_level_single_objective}\\
\toprule
Task & HEBO & BoFire & BT-L & RF & BT-U & CMA & TPE & Gryffin & Sobol & RS \\
\midrule
\endfirsthead
\multicolumn{11}{c}{\tablename~\thetable{} (continued)}\\
\toprule
Task & HEBO & BoFire & BT-L & RF & BT-U & CMA & TPE & Gryffin & Sobol & RS \\
\midrule
\endhead
\midrule
\multicolumn{11}{r}{Continued on next page}\\
\endfoot
\bottomrule
\endlastfoot
AgNP reaction rate & .065 & \textbf{.045} & .052 & .091 & .112 & .105 & .128 & .269 & .236 & .253 \\
OPV degradation & \textbf{.000} & \textbf{.000} & \textbf{.000} & .047 & \textbf{.000} & .048 & .055 & .093 & .093 & .098 \\
Mobile hydrogen evolution & .183 & .131 & .351 & .212 & .292 & \textbf{.120} & .364 & .589 & .572 & .641 \\
BRINE campaign 2 & .055 & .059 & .063 & .052 & .057 & .057 & \textbf{.049} & .134 & .133 & .152 \\
Amide yield, reaction 1 & \textbf{.003} & .008 & .009 & .009 & .008 & .009 & .006 & .017 & .020 & .021 \\
Printing compression modulus & \textbf{.019} & \textbf{.019} & \textbf{.019} & .210 & \textbf{.019} & .217 & .216 & .229 & .277 & .275 \\
N-benzylation impurity & \textbf{$<$.001} & .004 & .007 & .014 & .004 & .019 & .030 & .045 & .061 & .067 \\
SNAr E-factor & \textbf{$<$.001} & .002 & .002 & .011 & .002 & .007 & .022 & .083 & .108 & .108 \\
P3HT--CNT conductivity & .060 & .202 & \textbf{.046} & .154 & .454 & .154 & .206 & .237 & .247 & .240 \\
Printing toughness & \textbf{.061} & .215 & .202 & .090 & .251 & .116 & .096 & .181 & .193 & .172 \\
\midrule
Li-ion electrolyte conductivity & \textbf{$<$.001} & .015 & .013 & .014 & .018 & .011 & .027 & .085 & .065 & .074 \\
Amide yield, reaction 4 & \textbf{.001} & .001 & .007 & .068 & .006 & .028 & .100 & .293 & .269 & .330 \\
Amide yield, reaction 2 & \textbf{.002} & .055 & .062 & .039 & .036 & .047 & .033 & .105 & .105 & .149 \\
BRINE campaign 1 & \textbf{.085} & .092 & .093 & .103 & .092 & .091 & .110 & .160 & .203 & .205 \\
\end{longtable}
\end{landscape}

The tier sensitivity analysis repeated the within-task median ranking separately
for the ten Core and four Conditional tasks. The resulting mean ranks are shown
in Table~\ref{tab:tier_sensitivity_ranks}; their Spearman correlation is 0.794.

\begin{table}[h]
\centering
\caption{Mean within-task rank by oracle-admission tier. Lower is better.}
\label{tab:tier_sensitivity_ranks}
\small
\begin{tabular}{lrr}
\toprule
Algorithm & Core (10 tasks) & Conditional (4 tasks) \\
\midrule
HEBO & 2.00 & 1.00 \\
BoFire LogEI & 3.85 & 4.25 \\
BoTorch LogEI & 4.10 & 4.75 \\
RF/SMAC-like EI & 4.20 & 5.00 \\
BoTorch UCB & 4.75 & 3.75 \\
CMA-ES & 5.00 & 3.50 \\
Optuna TPE & 5.30 & 5.75 \\
Gryffin & 8.10 & 9.00 \\
Sobol Search & 8.60 & 8.25 \\
Random Search & 9.10 & 9.75 \\
\bottomrule
\end{tabular}
\end{table}

The single-objective omnibus comparison used a Friedman test with the 14 tasks
as blocks and the task-level median final regrets as observations. The
follow-up comparisons reported in the main text are two-sided Wilcoxon
signed-rank tests between HEBO and each alternative over the same 14 task
medians. The resulting family of nine $p$-values was adjusted by Holm's method.

\subsection{Optimization trajectories}

\begin{figure}[p]
\centering
\includegraphics[width=\textwidth]{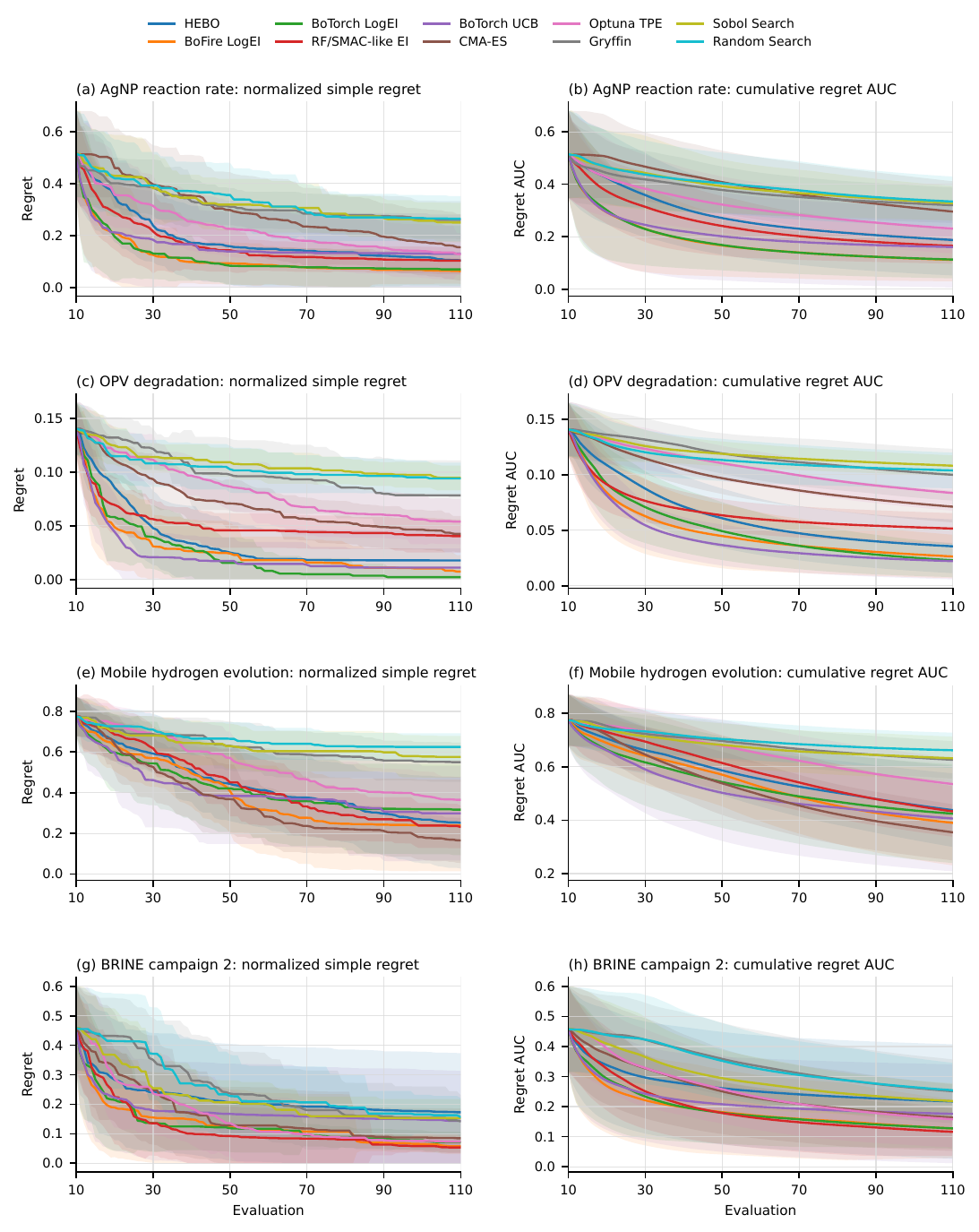}
\caption{Mean single-objective optimization trajectories (solid lines) with
one sample standard deviation across 20 trials (shaded regions) for AgNP
reaction rate, OPV degradation, mobile hydrogen evolution, and BRINE campaign
2. Left panels show normalized best-so-far simple regret; right panels show
running normalized regret AUC. Lower values are better.}
\label{fig:so_trajectories_1}
\end{figure}

\begin{figure}[p]
\centering
\includegraphics[width=\textwidth]{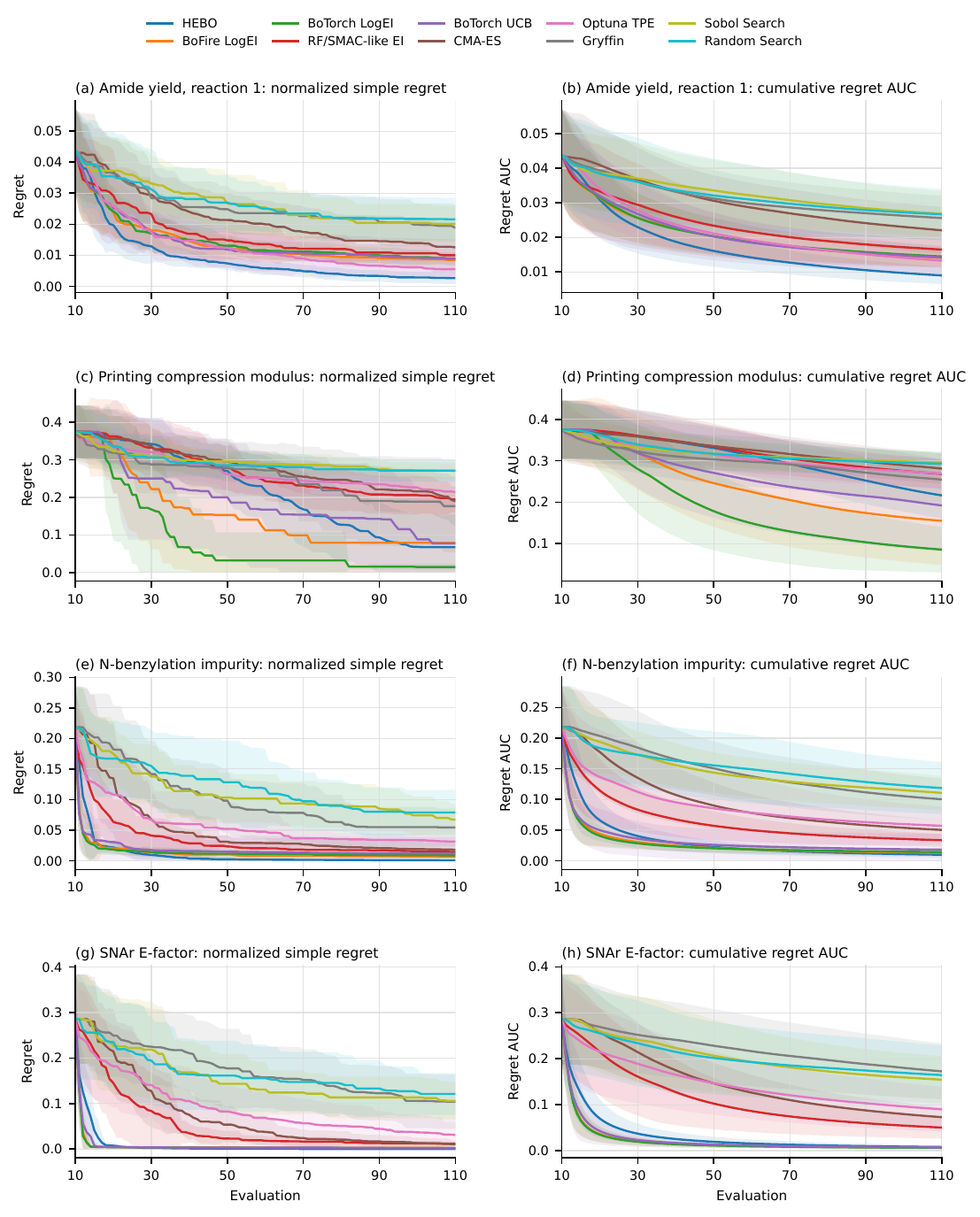}
\caption{Mean single-objective optimization trajectories (solid lines) with
one sample standard deviation across 20 trials (shaded regions) for amide yield
(reaction 1), printing compression modulus, N-benzylation impurity, and SNAr
E-factor. Left panels show normalized best-so-far simple regret; right panels
show running normalized regret AUC. Lower values are better.}
\label{fig:so_trajectories_2}
\end{figure}

\begin{figure}[p]
\centering
\includegraphics[width=\textwidth]{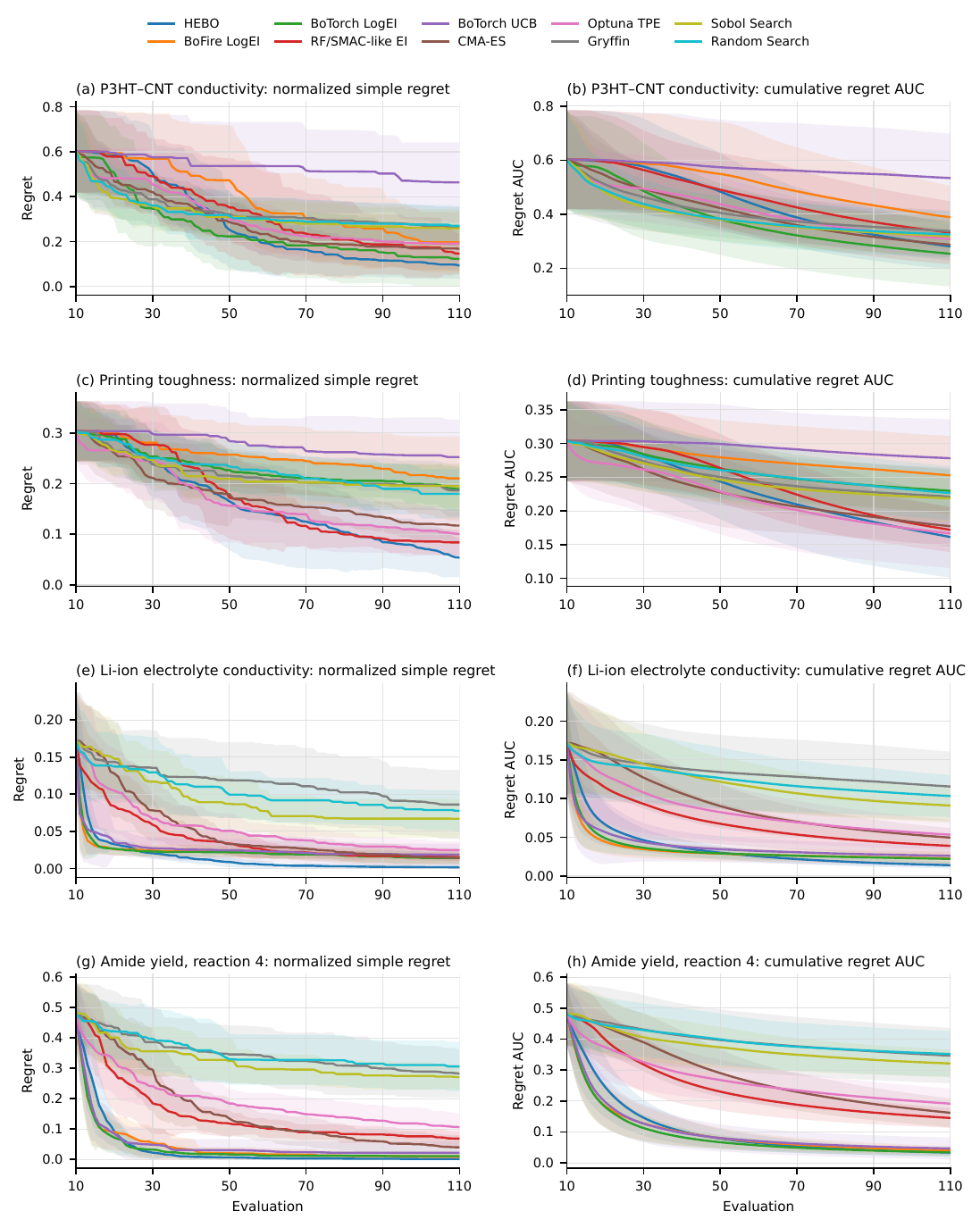}
\caption{Mean single-objective optimization trajectories (solid lines) with
one sample standard deviation across 20 trials (shaded regions) for P3HT--CNT
conductivity, printing toughness, Li-ion electrolyte conductivity, and amide
yield (reaction 4). Left panels show normalized best-so-far simple regret;
right panels show running normalized regret AUC. Lower values are better.}
\label{fig:so_trajectories_3}
\end{figure}

\begin{figure}[p]
\centering
\includegraphics[width=\textwidth]{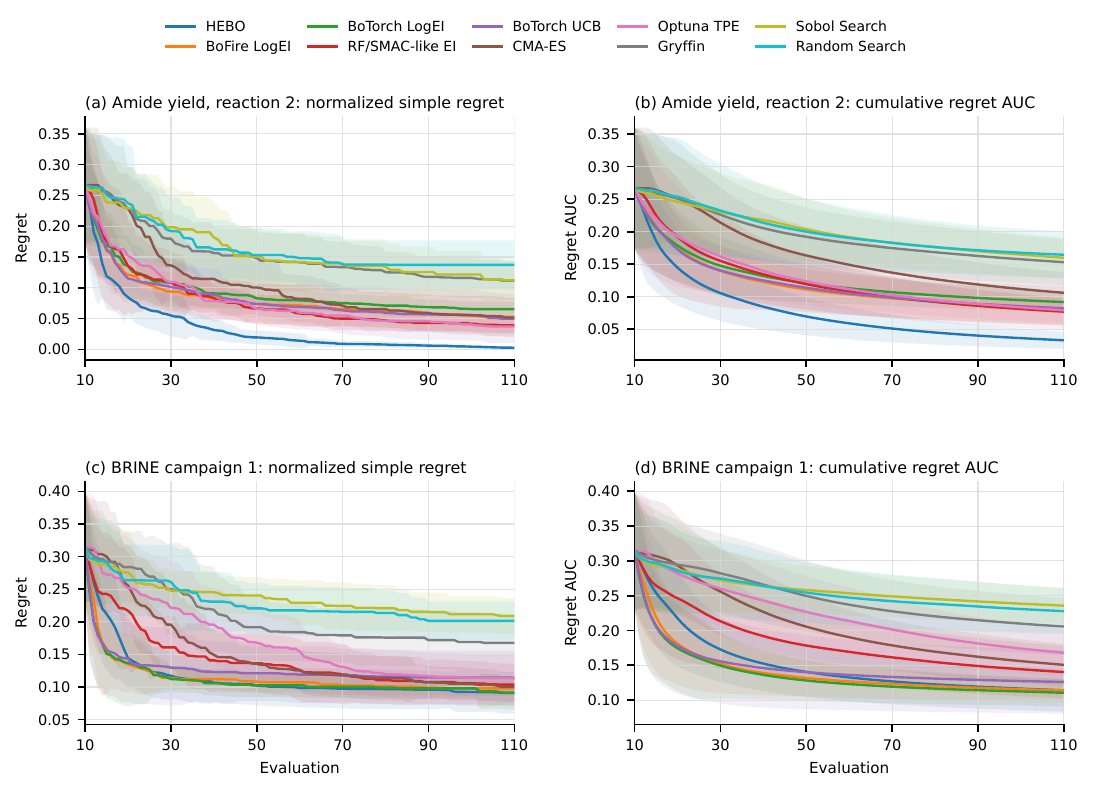}
\caption{Mean single-objective optimization trajectories (solid lines) with
one sample standard deviation across 20 trials (shaded regions) for amide yield
(reaction 2) and BRINE campaign 1. Left panels show normalized best-so-far
simple regret; right panels show running normalized regret AUC. Lower values
are better.}
\label{fig:so_trajectories_4}
\end{figure}

Figures~\ref{fig:so_trajectories_1}--\ref{fig:so_trajectories_4} show the
complete single-objective trajectories underlying the endpoint summaries in
the main text. For each task, the left panel is the normalized best-so-far
simple regret at evaluation $t$, and the right panel is the running normalized
trapezoidal AUC from evaluation 10 through $t$. At $t=10$, the latter is
defined by continuity as the regret at evaluation 10. Its value at $t=110$
therefore equals the normalized regret AUC reported in the main text. Solid
lines show means over the 20 trials, and shaded regions show one sample standard
deviation around the mean; the lower edge of each band is clipped at zero for
display. Lower values are better in all single-objective panels.

\begin{figure}[p]
\centering
\includegraphics[width=\textwidth]{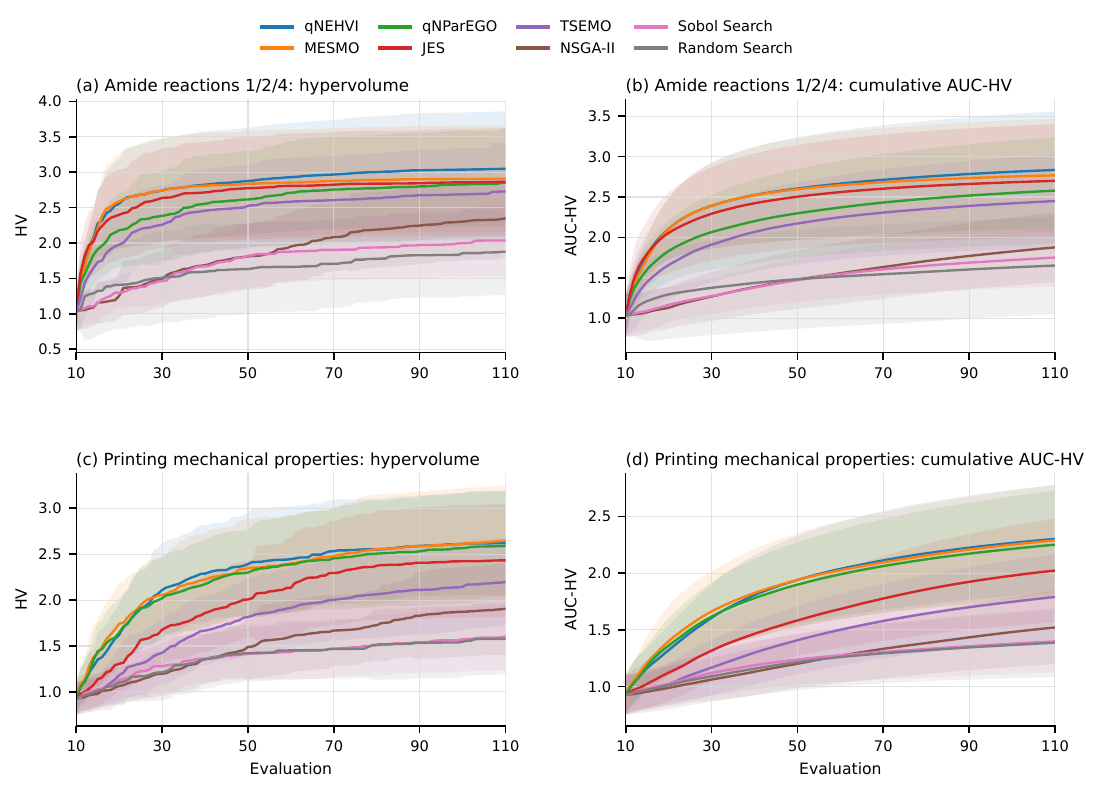}
\caption{Mean multi-objective optimization trajectories (solid lines) with one
sample standard deviation across 20 trials (shaded regions) for the combined
amide-reaction task and the printing-mechanical-properties task. Left panels
show hypervolume after trial-specific objective normalization; right panels
show running AUC-HV. Higher values are better.}
\label{fig:mo_trajectories}
\end{figure}

Figure~\ref{fig:mo_trajectories} reports the corresponding multi-objective
trajectories. Hypervolume at each budget was computed using the trial-specific
objective scaling and fixed reference point defined in the main text. Running
AUC-HV is the trapezoidal average from evaluation 10 through the current
evaluation, with its value at evaluation 10 defined by continuity. Solid lines
and shaded regions again denote the mean and one sample standard deviation over
20 trials. Higher values are better in all multi-objective panels.

\clearpage

\clearpage

\subsection{Preliminary Evaluation of Agentic Optimization}
\label{app:agentic}

To provide a preliminary comparison with agentic optimization, we evaluated
ScienceFlow using two language-model backends. We considered ScienceFlow-K,
which was supplied with human-readable task and variable semantics, and
ScienceFlow-M, which received only the information available to the
conventional optimizers. All experimental settings and evaluation procedures
were identical to those used in the main single-objective experiments.

\begin{table*}[t]
    \centering
    \caption{Aggregate single-objective performance of the conventional
    optimizers and the evaluated ScienceFlow configurations. Mean ranks are
    recomputed jointly over all 14 entries from task-level median final
    performance. Regret values are reported as mean $\pm$ standard deviation
    over 280 task--trial runs. Lower values are better for all metrics.}
    \label{tab:scienceflow_comparison}
    \small
    \setlength{\tabcolsep}{5pt}
    \begin{tabular}{lccc}
        \toprule
        Algorithm
        & \makecell{Mean task\\rank $\downarrow$}
        & \makecell{Final normalized\\simple regret $\downarrow$}
        & \makecell{Normalized regret\\AUC $\downarrow$} \\
        \midrule
        HEBO
        & \textbf{2.07}
        & $\mathbf{0.062 \pm 0.117}$
        & $0.126 \pm 0.148$ \\

        BoFire LogEI
        & 4.61
        & $0.075 \pm 0.114$
        & $0.125 \pm 0.145$ \\

        BoTorch LogEI
        & 4.86
        & $0.071 \pm 0.112$
        & $\mathbf{0.111 \pm 0.133}$ \\

        RF/SMAC-like EI
        & 5.07
        & $0.080 \pm 0.080$
        & $0.146 \pm 0.130$ \\

        CMA-ES
        & 5.29
        & $0.084 \pm 0.089$
        & $0.160 \pm 0.118$ \\

        BoTorch UCB
        & 5.46
        & $0.115 \pm 0.169$
        & $0.149 \pm 0.183$ \\

        Optuna TPE
        & 6.57
        & $0.105 \pm 0.108$
        & $0.172 \pm 0.142$ \\

        ScienceFlow-K (DeepSeek-V4-Flash)
        & 8.79
        & $0.190 \pm 0.198$
        & $0.225 \pm 0.205$ \\

        ScienceFlow-M (DeepSeek-V4-Flash)
        & 9.18
        & $0.163 \pm 0.156$
        & $0.218 \pm 0.172$ \\

        ScienceFlow-K (GLM-5.3-Flash)
        & 9.39
        & $0.214 \pm 0.235$
        & $0.247 \pm 0.231$ \\

        Gryffin
        & 10.29
        & $0.178 \pm 0.146$
        & $0.231 \pm 0.156$ \\

        Sobol Search
        & 10.57
        & $0.189 \pm 0.144$
        & $0.230 \pm 0.151$ \\

        ScienceFlow-M (GLM-5.3-Flash)
        & 11.21
        & $0.214 \pm 0.202$
        & $0.256 \pm 0.207$ \\

        Random Search
        & 11.64
        & $0.200 \pm 0.154$
        & $0.240 \pm 0.161$ \\
        \bottomrule
    \end{tabular}
\end{table*}

The comparison yields a predominantly negative assessment of the current
agentic approach. All four ScienceFlow configurations rank below seven
dedicated numerical optimizers, with the highest-ranked configuration
placing only eighth among the 14 evaluated entries. The best aggregate final
regret and regret AUC obtained by ScienceFlow are 0.163 and 0.218,
respectively, compared with 0.062 and 0.111 for the best conventional
results. The ScienceFlow configurations also exhibit substantial variation
across tasks and trials. Although isolated runs may produce competitive
solutions, these successes do not translate into consistently competitive
optimization performance. Taken together, these results suggest that current agentic optimization
approaches may not yet be competitive with established numerical
optimization algorithms on this benchmark.

\section{Reliability-analysis details}
\label{app:d}

\subsection{Independent oracle retraining}

For seeds 2026, 2027, and 2028, Python's random module, NumPy,
\texttt{PYTHONHASHSEED}, and the random-forest and extra-trees model seeds were
set to that value. Six tasks were then refitted independently with the same
AutoGluon protocol described in Appendix~\ref{app:b}. Each seed-specific model
was evaluated with the same 10 algorithms, 20 trials, common initialization,
and post-initialization optimizer streams, yielding 3,600 complete runs and
396,000 oracle evaluations. Rankings were calculated within task--trial blocks
and then averaged over the six tasks.

\begin{table}[h]
\centering
\caption{Agreement between the ten-algorithm rankings obtained from independently retrained oracles.}
\label{tab:oracle_seed_rank_agreement}
\small
\begin{tabular}{ccrrc}
\toprule
Seed A & Seed B & Spearman $\rho$ & Kendall $\tau_b$ & Top-three overlap \\
\midrule
2026 & 2027 & 0.964 & 0.911 & 3 \\
2026 & 2028 & 1.000 & 1.000 & 3 \\
2027 & 2028 & 0.964 & 0.911 & 3 \\
\bottomrule
\end{tabular}
\end{table}

Numerical agreement was assessed on identical response-independent candidates
from the Random and Sobol histories. For each task and seed pair, Pearson and
Spearman correlations and a mean absolute error normalized by the pooled
prediction range were computed. Candidate identity was 1.0 in every comparison.
Table~\ref{tab:oracle_prediction_agreement} aggregates the three seed pairs for
each task.

\begin{table*}[h]
\centering
\caption{Prediction agreement across independent oracle fits. NMAE is mean absolute error divided by the task prediction range.}
\label{tab:oracle_prediction_agreement}
\small
\begin{tabular}{lrrrr}
\toprule
Task & Mean Pearson & Minimum Spearman & Mean NMAE & Maximum NMAE \\
\midrule
Mobile hydrogen evolution & 0.9863 & 0.9760 & 0.0155 & 0.0179 \\
AgNP reaction rate & 0.9978 & 0.9833 & 0.0088 & 0.0123 \\
BRINE campaign 2 & 0.9959 & 0.9951 & 0.0111 & 0.0118 \\
Printing compression modulus & 0.9976 & 0.9963 & 0.0103 & 0.0115 \\
Amide yield, reaction 1 & 0.9986 & 0.9976 & 0.0086 & 0.0090 \\
OPV degradation & 0.9989 & 0.9982 & 0.0046 & 0.0050 \\
\bottomrule
\end{tabular}
\end{table*}

\subsection{Measured-table replay}

Measured replay operates on a finite pool of unique experimental conditions.
Rows with identical model inputs were collapsed to one condition by averaging
their responses; the replicate count and response standard deviation remain in
the pool. Each task--trial pair begins from ten frozen measured rows shared by
all algorithms. At each later step, an optimizer proposes a condition in the
same external domain used by the continuous oracle. For numerical variable $j$,
the component matching distance between proposal $x$ and measured row $x'$ is
\[
\delta_j(x,x')=\frac{|x_j-x'_j|}{u_j-l_j},
\]
and for a categorical variable it is the zero--one mismatch indicator. The row
distance is the unweighted mean $D(x,x')=p^{-1}\sum_{j=1}^p\delta_j(x,x')$.
The closest unused row is selected; numerical ties are ordered by the stored
opaque row identifier. That row is removed from the pool, and the optimizer is
told the matched measured condition and measured response rather than its
original proposal. Consequently, no measured row can be evaluated twice in a
run.

\begin{table}[h]
\centering
\caption{Measured replay pool construction. The 60-evaluation comparison uses
all six tasks; the 110-evaluation comparison uses the two pools with at least
110 unique conditions after initialization.}
\label{tab:replay_pools}
\small
\begin{tabular}{lrrrcc}
\toprule
Task & Valid rows & Unique conditions & Collapsed rows & 60 & 110 \\
\midrule
AgNP reaction rate & 144 & 130 & 14 & Yes & No \\
OPV degradation & 2322 & 2272 & 50 & Yes & Yes \\
Mobile hydrogen evolution & 595 & 521 & 74 & Yes & Yes \\
BRINE campaign 2 & 121 & 121 & 0 & Yes & No \\
Amide yield, reaction 1 & 117 & 116 & 1 & Yes & No \\
Printing compression modulus & 143 & 143 & 0 & Yes & No \\
\bottomrule
\end{tabular}
\end{table}

Matching diagnostics were calculated over the post-initialization evaluations
of all ten algorithms and 20 trials. Table~\ref{tab:replay_matching} reports the
median of run-level median distances, the mean run-level 95th percentile, the
largest observed distance, and the fraction of matches exceeding 0.25.

\begin{table*}[h]
\centering
\caption{Measured-table replay matching distances by task and total budget.}
\label{tab:replay_matching}
\small
\begin{tabular}{lrrrrr}
\toprule
Task & Budget & Median run median & Mean run P95 & Maximum & Fraction $>0.25$ \\
\midrule
AgNP reaction rate & 60 & 0.141 & 0.239 & 0.452 & 0.068 \\
Amide yield, reaction 1 & 60 & 0.231 & 0.338 & 0.510 & 0.375 \\
BRINE campaign 2 & 60 & 0.041 & 0.084 & 0.183 & 0.000 \\
Mobile hydrogen evolution & 60 & 0.068 & 0.103 & 0.166 & 0.000 \\
OPV degradation & 60 & 0.016 & 0.056 & 0.136 & 0.000 \\
Printing compression modulus & 60 & 0.109 & 0.191 & 0.301 & 0.024 \\
Mobile hydrogen evolution & 110 & 0.080 & 0.122 & 0.196 & 0.000 \\
OPV degradation & 110 & 0.020 & 0.074 & 0.162 & 0.000 \\
\bottomrule
\end{tabular}
\end{table*}

At 60 evaluations, 1,200 runs covered all six replay tasks and ten algorithms.
At 110 evaluations, 400 runs covered OPV degradation and mobile hydrogen
evolution. The agreement statistics in Table~\ref{tab:replay_rank_agreement}
compare the three-seed oracle aggregate with measured replay at the same budget.
The two rows have different task sets and therefore should be interpreted as
separate comparisons rather than as a controlled estimate of the effect of
budget alone.

\begin{table}[h]
\centering
\caption{Rank agreement between learned-oracle optimization and measured-table replay.}
\label{tab:replay_rank_agreement}
\small
\begin{tabular}{rrrrr}
\toprule
Budget & Tasks & Spearman $\rho$ & Kendall $\tau_b$ & Top-three overlap \\
\midrule
60 & 6 & 0.758 & 0.556 & 2 \\
110 & 2 & 0.455 & 0.333 & 2 \\
\bottomrule
\end{tabular}
\end{table}

\end{document}